\documentclass{elsarticle}

\pdfoutput=1

\begin{document}

\title{Cryogenic SiPM arrays for the DUNE photon detection system}
\author[a]{A.~Falcone\corref{cor}}
\author[c]{A.~Andreani}
\author[b]{S.~Bertolucci}
\author[a]{C.~Brizzolari}
\author[d]{N.~Buckanam}
\author[f]{M.~Capasso}
\author[a]{C.~Cattadori}
\author[a]{P.~Carniti}
\author[c]{M.~Citterio}
\author[e]{K.~Francis}
\author[c]{N.~Gallice}
\author[f]{A.~Gola}
\author[a]{C.~Gotti}
\author[b]{I.~Lax}
\author[g]{P.~Litrico}
\author[f]{A.~Mazzi}
\author[a]{M.~Mellinato}
\author[b]{A.~Montanari}
\author[b]{L.~Patrizii}
\author[b]{L.~Pasqualini}
\author[a]{G.~Pessina}
\author[b]{M.~Pozzato}
\author[c]{S.~Riboldi}
\author[c]{P.~Sala}
\author[b]{G.~Sirri}
\author[a,b]{M.~Tenti}
\author[a]{F.~Terranova}
\author[a]{M.~Torti}
\author[b]{R.~Travaglini}
\author[d]{D.~Warner}
\author[d]{R.~Wilson}
\author[a]{V.~Zutshi}

\address[a]{INFN Milano Bicocca and University of Milano Bicocca, Department of Physics, Milano, Italy}
\address[b]{INFN Bologna and University of Bologna, Department of Physics, Bologna, Italy}
\address[c]{INFN Milano and University of Milano, Department of Physics, Milano, Italy}
\address[d]{Colorado State University, Fort Collins, CO, USA}
\address[e]{Northern Illinois University, Department of Physics, DeKalb, IL, USA}
\address[f]{Fondazione Bruno Kessler, Trento, Italy}
\address[g]{INFN, Laboratori Nazionali del Sud (LNS), Catania, Italy}

\cortext[cor]{Corresponding author}

\begin{abstract}
In this paper we report on the characterization of SiPM tiles developed for the R\&D on the DUNE Photon Detection System. The tiles were produced by Fondazione Bruno Kessler (FBK) employing NUV-HD-SF SiPMs.  Special emphasis is given on cryo-reliability 
of the sensors, i.e. the stability of electric and mechanical properties after thermal cycles at room and 77~K temperature. The characterization includes the determination of the I-V curve, a high sensitivity measurement of Dark Count Rate at different overvoltages, and correlated noise. The single p.e. sensitivity is measured as a function of the number of sensors connected to a single electronic channel, after amplification at 77~K using a dedicated cold amplifier.
\end{abstract}

\maketitle

\section{Introduction}
\begin{figure}[htbp]
\begin{center}
    \includegraphics[keepaspectratio=true,scale=0.65]{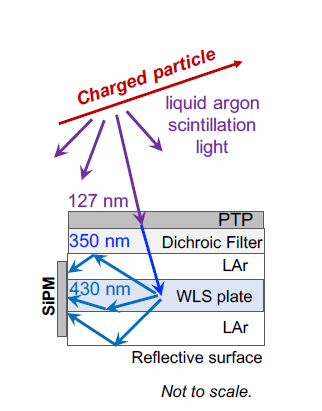}
\end{center}
    \caption{Working principle of an X-ARAPUCA cell.}
    \label{fig:arapuca}
\end{figure}

UV photon detection in liquefied noble gases plays a prominent role in several Neutrino and Dark Matter experiments, whose needs have fostered the development of cryogenic Silicon Photomultipliers (SiPMs). Among the experiments that employ a Photon Detection System (PDS) for the scintillation light of liquid argon, DUNE~\cite{DUNE} poses major challenges in terms of scalability to large volumes (40~kton per module) and long-term reliability. In particular, the first DUNE module is based on a compact Anode Plane Assembly \cite{Abi:2018alz} where the PDS must be located. Space constraints thus highly favor the use of SiPM arrays, which must be coupled to a system for light trapping and transport. The first DUNE module will use the X-Arapuca as light trapping  system. The  ARAPUCA \cite{Machado:2016jqe} is a light trap that captures wavelength-shifted photons inside a box with highly reflective internal surfaces until they
are eventually detected by SiPMs (see Fig. \ref{fig:arapuca}). An X-ARAPUCA~\cite{Machado:2018rfb} cell is based on a 10.0$\times$7.8~cm$^2$ dychroic filter coated with p-Terphenyl (PTP). The argon scintillation photons at 128~nm impinging on the PTP layer are shifted to 350~nm. The photons emitted toward the cell enter the filter, which has a cutoff  wavelength of about 400~nm. A Wavelength Shifting (WLS) bar with an emission wavelength
larger then the transmission range of the filter is installed inside the cell, while the inner lower surface is covered with a reflective layer. Wavelength shifted photons from this plate are either transported along the WLS plate to the photosensors via total internal reflection, or if they escape the plate, may be captured within the dichroic filter. Arrays of SiPMs are positioned along the lateral walls of the cell to detect the photons. %An X-Arapuca cell is read out by SiPMs located along the lateral walls of the cell. 
The signals of multiple SiPMs are summed and amplified at cold, i.e. the SiPMs are operated in ganging mode. In the course of the R\&D for the X-Arapuca, several SiPM technologies have been tested in view of the installation in ProtoDUNE-SP in 2021 (Run II)~\cite{Abi:2017aow}. In this paper, we report the results of the characterization of SiPM arrays produced by Fondazione Bruno Kessler (FBK) in 2018 and based on the NUV-HD-SF technology. NUV-HD-SF has been one of the first technologies tested for the new X-Arapuca design because of its excellent performance at cryogenic temperatures and ideal matching with the emission spectrum of the WLS bars (Eljen EJ-286). Special emphasis has been given to cryo-reliability, thermal tests and mechanical robustness of the packaging (Sec.~\ref{sec:thermal_test}) and the performance in ganged mode (Sec.~\ref{sec:ganging}). The characterization of the tiles in term I-V curves, dark count rate and correlated noise are summarized in Sec.~\ref{sec:ivcurve} and Sec.~\ref{sec:DCR}, respectively.          

\section{Experimental Setup}
\label{sec:setup}

\subsection{Silicon Photomultiplier arrays}
\begin{figure}[htbp]
    \centering
    \includegraphics[keepaspectratio=true,scale=0.75]{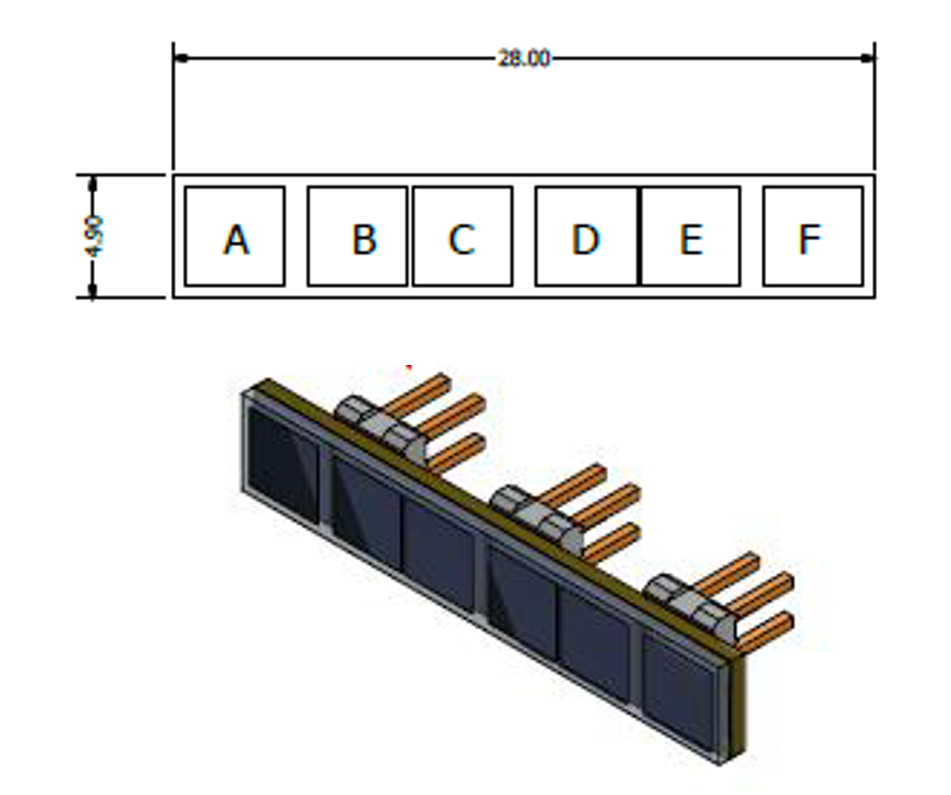}
    \caption{Schematic of a tile made of six 4$\times$4~mm$^2$ NUV-HD-SF photosensors.}
    \label{fig:Tile}
\end{figure}

FBK has been developing technologies for SiPMs since 2006~\cite{Piemonte_2006} and in 2016 delivered a technology (NUV-HD)~\cite{gola_2016,Gola:2019idb}  tuned for the near UV, that has been successfully tested at room and cryogenic temperatures~\cite{Acerbi:2016ikf}. In particular the NUV-HD technology is based on p-on-n junctions. Microcells are separated by deep trenches that provide electrical isolation. Trenches are filled with silicon dioxide. Due to the different refractive index, these trenches also allow for partial optical isolation between microcells. The NUV-HD technology is available both in the "standard" and "low" field option. The standard field (SF) sensors, in particular, have been produced in a broad range of microcell sizes (from 15 to 40~$\mu$m) and form factors. The tiles produced for the DUNE R\&D are assembled from 4$\times$4~mm$^2$ NUV-HD-SF SiPMs. The SiPMs have a cell-pitch size of 40~$\mu$m and a microcell fill factor of 83\%.
They are arranged in 6~unit tiles (Fig~\ref{fig:Tile}). The SiPMs were mounted on a printed circuit board and connected through wire bondings. Each SiPM can be read out individually by pin pairs. The front side is protected with a layer of epoxy resin.

\subsection{Test facilities and equipment}
The measurements performed on the tiles were carried on using a low noise cryogenic facility complemented by ancillary equipment for I-V curve and thermal tests. 

The I-V curves were acquired re-optimizing a test facility for the cryogenic characterization of diodes and transistors. It is based on a Keithley 4200A-SCS semiconductor analyzer which is able to deliver bias voltage in the 0-200~V range recording currents with a precision $<$1~pA.
The analyser was connected to the anode and cathode of all SiPMs belonging to the same tile in order to characterize the sensors separately. A dedicated board able to host up to three tiles was installed at the cold end of a hollow 80 cm-long steel bar. The shaft can be manually positioned inside a liquid nitrogen dewar and a steel shield protects the board from a sudden direct contact with the liquid nitrogen (LN) during the cooling down. Temperature is monitored inside the shield in thermal contact with the board by a PT100  platinum resistance thermometer. The LN penetrates the shield and thermalizes the board at 77~K in less then 5~minutes. The warm end of the bar is a metallic box that hosts the BNC connectors to read the current drawn by the SiPMs, the PT100 and an optical fiber. The fiber is employed to illuminate the sensors with a LED (see below)  with tunable intensity during the test in reverse bias, to allow a precise identification of the breakdown voltage. The signal from the SiPM is brought to the analyzer and recorded.
The precision on the current achieved with this apparatus, being $\sim$1~pA, is sufficient to perform a characterization of the SiPMs in term of current response to applied voltage (I-V curve) both in forward and reverse bias.

\begin{figure}[htbp]
    \centering
    \includegraphics[keepaspectratio=true,scale=0.05]{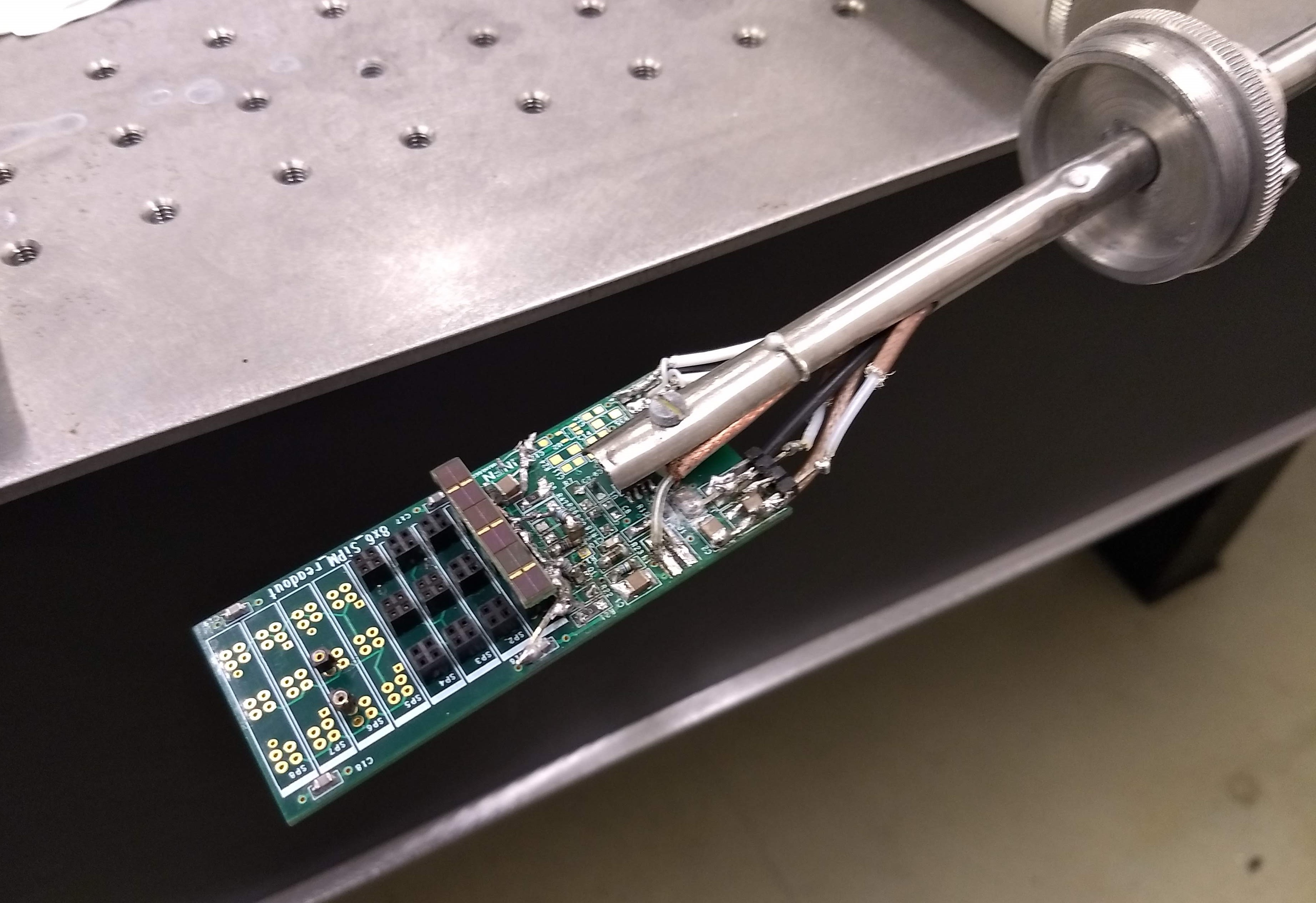}
    \caption{Cold amplification board (CAB) used to host the cold amplification stage and the SiPM tiles.}
    \label{fig:Board}
\end{figure}

The main characterization facility is based on a cold amplification board (CAB) hosting the SiPM tiles and a trans-impedance amplifier (TIA) tuned for cryogenic applications. The CAB is shown in Fig.~\ref{fig:Board}.
For the NUV-HD-SF the SiPMs were read by means of a two-stage amplifier. The first amplification stage is located in the CAB  and it is hence in thermal contact with the SiPMs. It is composed by a silicon-germanium heterojunction bipolar transistor (Infineon BFP640) and a fully differential operational amplifier (Texas Instruments THS4131). The configuration chosen for these tests is a low dynamic range version of the cold amplifier developed in~\cite{ampl}. This version is well suited for dark count and correlated noise measurements. The cold setup is similar to the setup used for the I-V curve and employs the same dewar. In the warm end of the hollow shaft, a metallic box hosts the second stage amplifier. 
This amplification stage is based on an OP27 (Texas Instruments) OpAmp, with a bandwidth of 300~kHz and gain 23.5 and it  is used to convert the differential signal to single ended one.

This test facility is also equipped with an illumination system.
The light is generated by a CAEN SP5601 LED pulser, which provides a 8~ns width pulse at $\lambda$~=~400~nm. It is brought to the SiPMs by an optical fiber, that runs inside the hollow bar connecting the two amplification stages together with signal and bias cables.

During data taking, signal from the second stage amplifier is sent to a Rohde-Schwarz RTO 1044 oscilloscope. The oscilloscope digitizes the signal waveform with a maximum sampling rate of 20~Gs/s  and 16-bit resolution.
Most of the runs comprise 10$^4$ waveforms. Each waveform records the signal for 4~$\mu$s at 10~GS/s.    

\section{Measurements and results}
Four NUV-HD-SF tiles were characterized both at room and LN (77~K) temperature using the test facilities mentioned above.

\subsection{Thermal tests}
\label{sec:thermal_test}

\begin{figure}[htbp]
    \centering
    \includegraphics[keepaspectratio=true,scale=0.25]{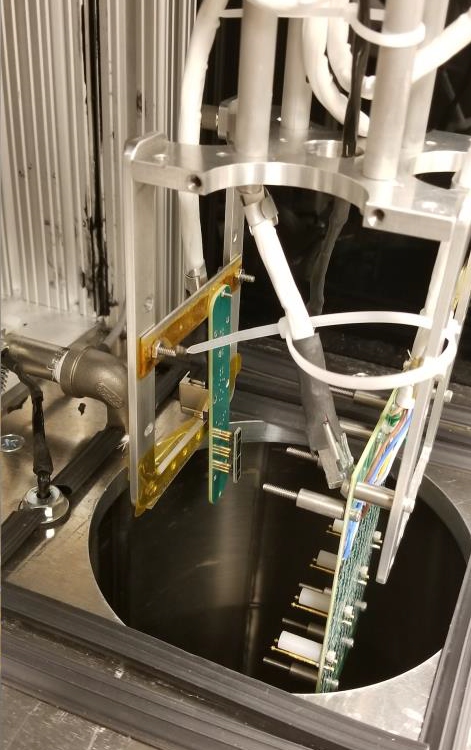}
    \caption{A NUV-HD-SF tile installed in the Colorado State University thermal test facility.}
    \label{fig:thermal_test}
\end{figure}

Cryo-reliability is a key issue in the design of the DUNE PDS. Thermal gradients stress the SiPMs at the level of detector, bonding and packaging.
SiPMs are arrays of  single-photon
avalanche diodes, each  integrated with a passive-quenching resistor, connected in parallel to common anode and cathode. As a consequence, operation at cryogenic temperatures changes most of the electric parameters of the SiPM and, in particular, the breakdown voltage and the quenching resistance. An ideal cryogenic photosensor is therefore a device where these changes are fully reversible over many thermal cycles. We tested the thermal robustness of the NUV-HD-SF in the wire bond epoxy package of the DUNE tiles in two datasets. The four tiles were cycled from 300~K to 77~K using a mechanically controlled test-stand at Colorado State University (Fig.~\ref{fig:thermal_test}). The response of the SiPMs at room temperature  after a cycle (gain, dark count and cross-talk) was checked as a function of the number of cycles and no degradation has been observed after $>20$~cycles. These measurements testify the mechanical integrity of the package and the bonding after cool-down and warm up cycles, and the reversibility of the electric behaviour of the device once it is operated back at room temperature. These results are corroborated in a direct manner by measuring the SiPM parameters at 77~K after several thermal cycles, as described below.

\subsection{I-V curve analysis}
\label{sec:ivcurve}
We used the system based on the 4200A-SCS analyzer (see Sec.~\ref{sec:setup}) to characterize the NUV-HD-SF in forward and reverse bias both at room and at 77~K. For an ideal cryogenic device we expect a drop of the breakdown voltage at 77~K. This is due to the increased mean free path between scattering of the carriers that are drifting in the high-field region. The longer mean free path increases the acceleration of the carriers at a given electric field and, therefore, breakdown can be achieved at a lower bias. For the case of NUV-HD-SF devices produced for DUNE, the breakdown voltage was measured biasing the devices in reverse mode and computing the maximum of $I^{-1}dI/dV$.  

\begin{figure}[htbp]
    \centering
    \includegraphics[keepaspectratio=true,scale=0.4]{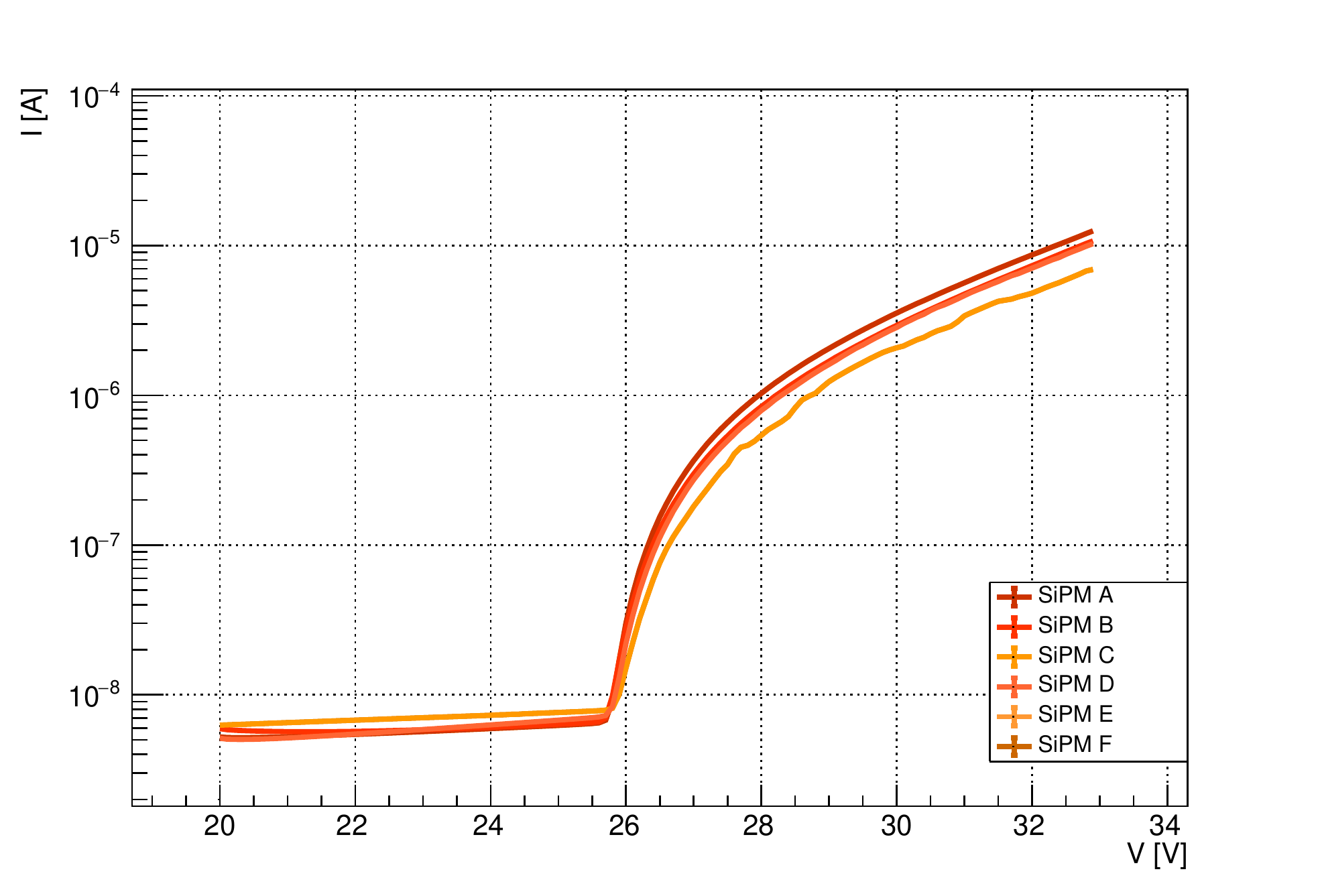}
    \includegraphics[keepaspectratio=true,scale=0.4]{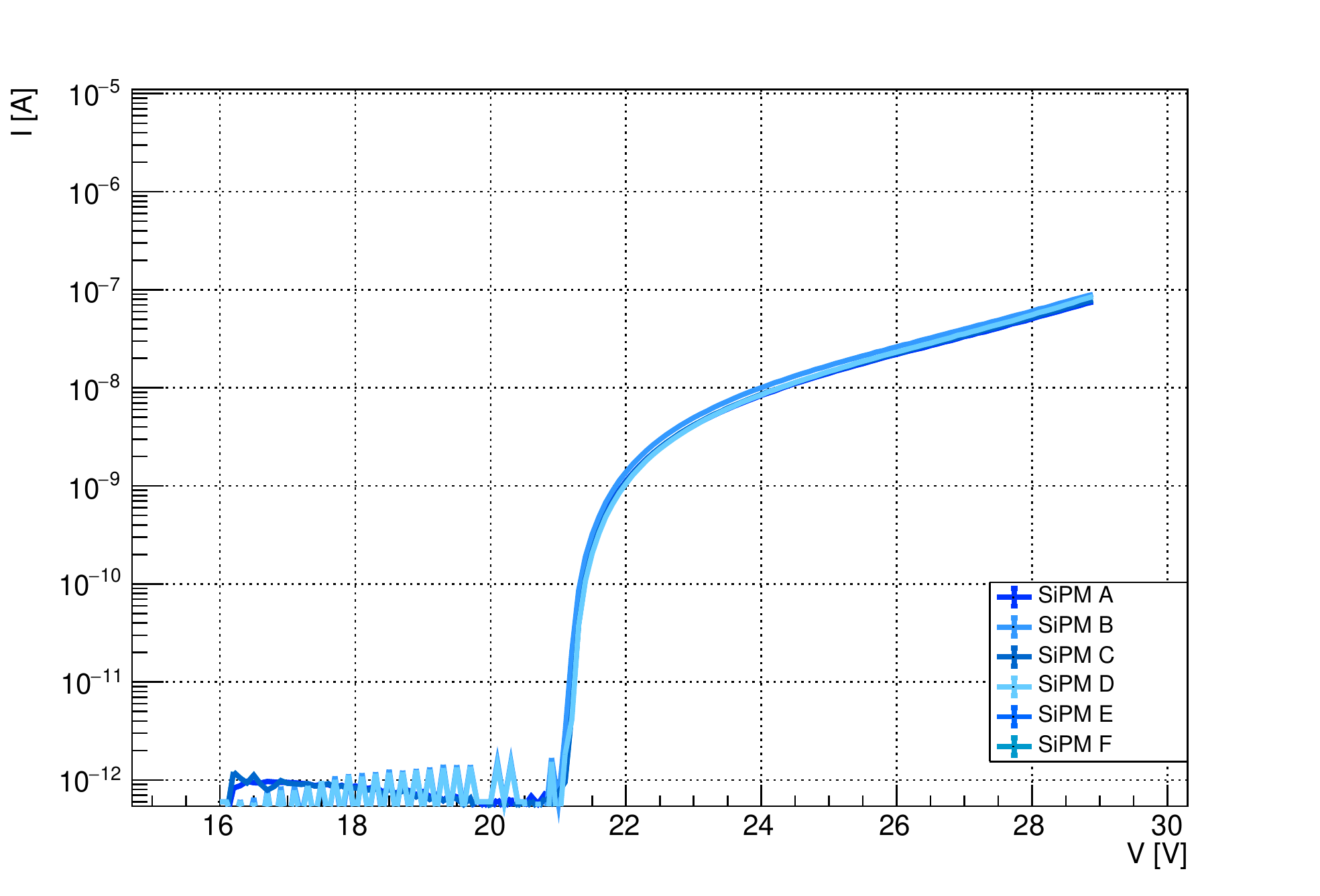}
    \caption{I-V curve of the SiPMs in one tile, biased in reverse mode, at room temperature (top) and T~=~77~K (bottom).}
    \label{fig:Reverse}
\end{figure}

%With the SiPM biased in reverse mode it is possible to measure the breakdown voltage of the devices, i.e. the voltage when the diode enter in avalanche mode. It corresponds to the maximum point of the second derivative of the I-V curve.
The measurement is made on each SiPM individually, in order to test the uniformity of this parameter both at room temperature and at 77~K.
The mean breakdown voltage among the various SiPMs, at room temperature, is V$_{bd}$~=~25.95~V with a standard deviation of 0.06~V, and drops to 21.19~V, with a standard deviation of 0.07~V, at 77~K. The average voltage drop is consistent with previous measurements of NUV-HD-SF~\cite{Acerbi:2016ikf,Gola:2019idb} at both temperatures and, in addition, $V_{bd}$ is very uniform among the devices.  The overvoltage (OV) working range is wide  (Fig.\ref{fig:Reverse}) and exceeds the range tested with the analyzer. For one of the SiPMs, the range was extended and the the occurrence of the second breakdown, which is commonly due to divergent correlated noise, was observed at $\sim$32~V, corresponding to a  maximum overvoltage range for the NUV-HD-SF at 77~K of +10~V.

\begin{figure}[htbp]
    \centering
    \includegraphics[keepaspectratio=true,scale=0.3]{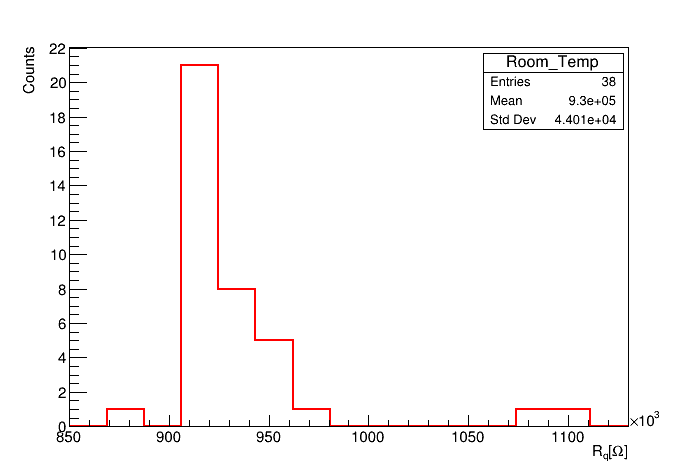}
    \includegraphics[keepaspectratio=true,scale=0.3]{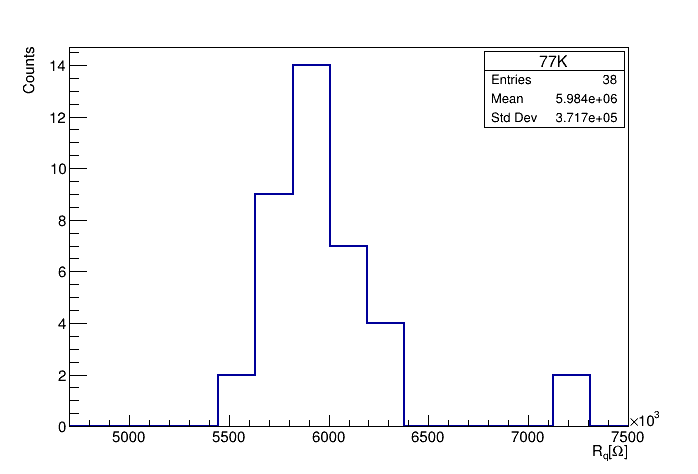}
    \caption{Quenching resistance distribution at room (top) and liquid nitrogen (bottom) temperature.}
    \label{fig:Resistance}
\end{figure}

Measurements of the I-V curves at forward bias provide a direct determination of the quenching resistance of the devices and its thermal coefficient. NUV-HD-SF SiPMs employ polysilicon resistors. Unlike metal resistors, they offer a higher range of tunability in the design of the sensors but also a large thermal coefficient.   
At 77~K (300~K), the quenching resistances $R_q$ (see Fig.~\ref{fig:Resistance}) is 5.9~M$\Omega$ (0.93~M$\Omega$) with a standard deviation of 370~k$\Omega$ (44~k$\Omega$). The standard deviation of the sample is dominated by two outliers at $R_{q}^{300K}~\simeq$~1~M$\Omega$, corresponding to SiPMs belonging to a different production batch.

\subsection{Dark and correlated noise}
\label{sec:DCR}
Extremely low values of dark count noise are one of the main strengths of cryogenic SiPMs and are a crucial parameter for most applications in liquid argon and xenon. Dark Count Rates (DCR) were measured for the NUV-HD-SF tiles at 77~K using the CAB-based facility described above. For these measurements the facility was moved in a dedicated dark room at the Dep.~of~Physics in Milano-Bicocca. A sensitivity to rates of about 100~mHz/mm$^2$ at 0.5~p.e. threshold was achieved by reducing environmental light and noise and filtering the output of the power supply that biases the SiPMs.  
The DCR is measured acquiring waveform without light pulse, with the trigger level at 0.5 photoelectrons. The results for one of the SiPMs are summarized in Tab.~\ref{tab:DCR}. DCR uniformity at single device level was tested for 6 SiPMs at +3 V of overvoltage and a threshold of 0.75 p.e. resulting in a mean DCR of 117~mHz/mm$^2$ and a standard deviation of 13~mHz/mm$^2$. 

\begin{table}
    \centering
    \begin{tabular}{|c|c|}
    \hline
    OV [V] & Counts/mm$^2$ [Hz]  \\
    \hline\hline
    3 & 0.135 \\
    \hline
    4 & 0.229 \\
    \hline
    5 & 0.520 \\
    \hline
    \end{tabular}
    \caption{Dark noise rate of a NUV-HD-SF SiPM as a function of overvoltage for a 0.5~p.e. threshold, at T~=~77~K.}
    \label{tab:DCR}
\end{table}

\begin{figure}[htbp]
    \centering
    \includegraphics[keepaspectratio=true,scale=0.4]{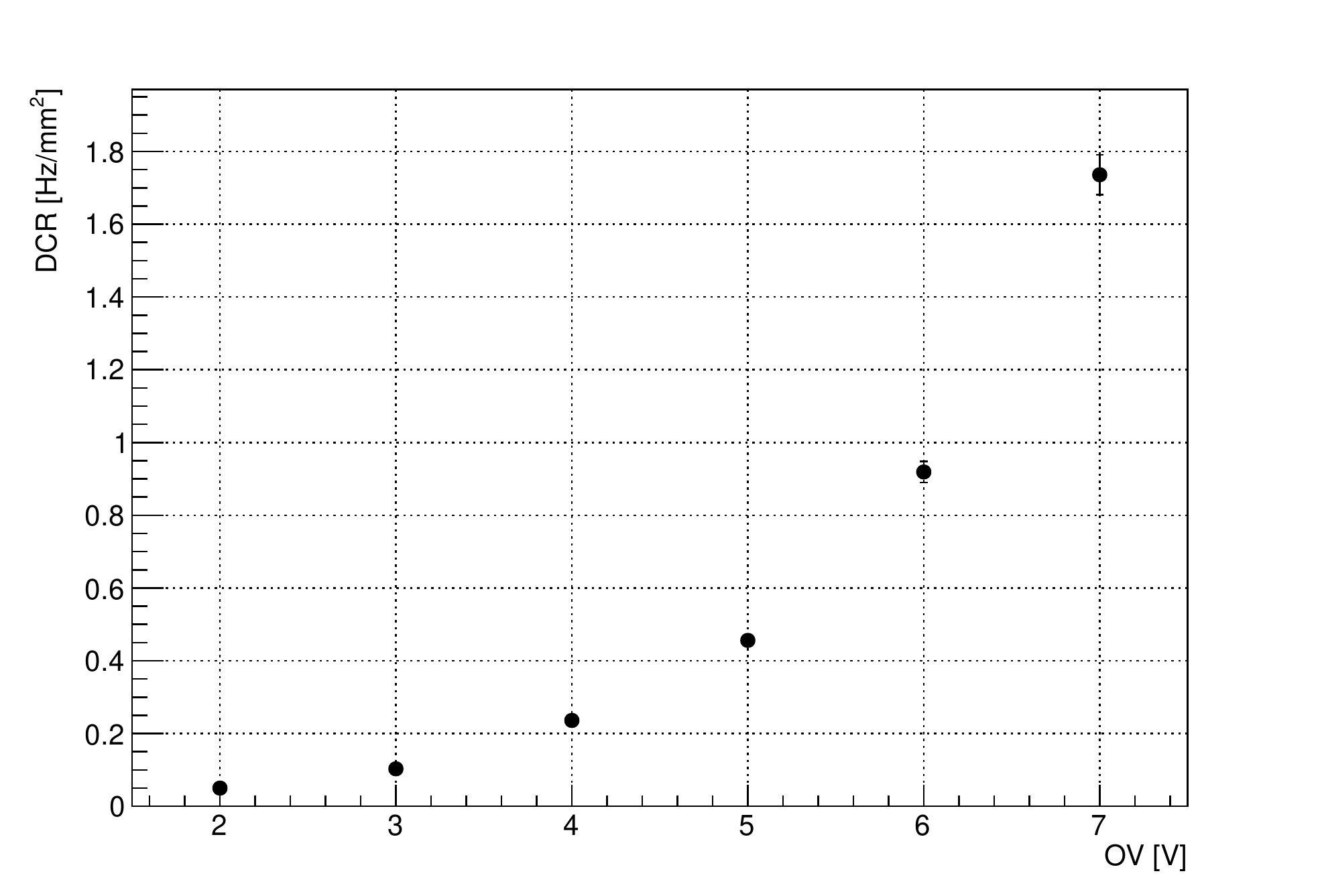}
    \caption{Dark noise rate as a single NUV-HD-SF as function of overvoltage for 0.75~p.e. threshold, at T~=~77~K.}
    \label{fig:DCRvsOV}
\end{figure}

\begin{figure}[htbp]
    \centering
    \includegraphics[keepaspectratio=true,scale=0.5]{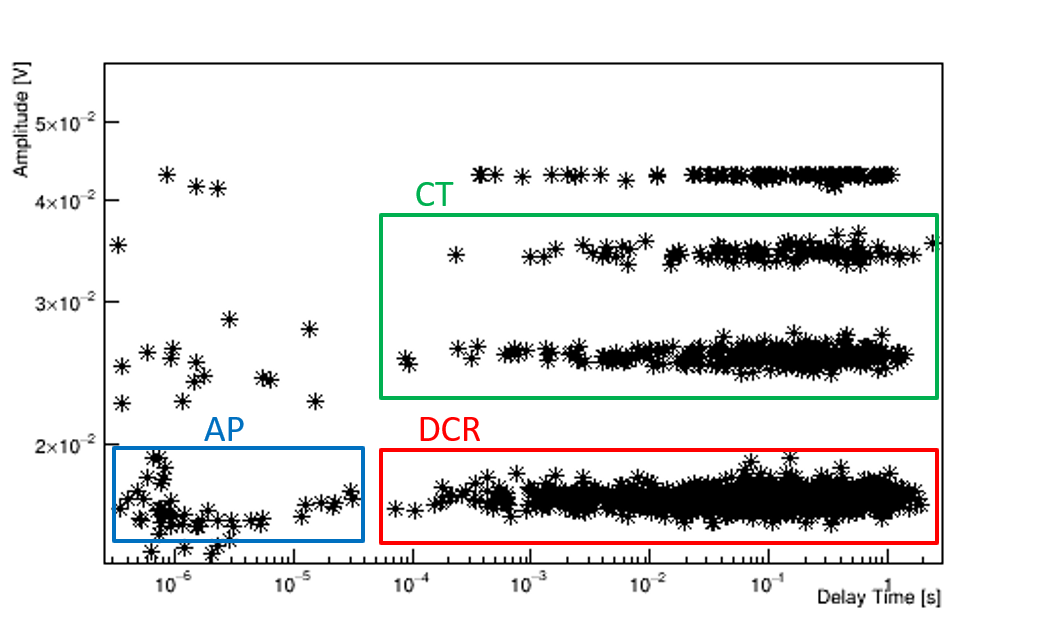}
    \caption{Distribution of delay time versus amplitude in dark of a single SiPM at T~=~77~K, with +4~OV bias. Primary dark count rate (DCR), direct crosstalks (CT) and afterpulses (AP) are shown in the red, green and blu boxes, respectively. }
    \label{fig:dark}
\end{figure}

Correlated noise in SiPMs can be measured by recording the time distance of signal peaks in standard DCR measurements. The 2-dim distribution of signal amplitudes versus time distance is shown in Fig.~\ref{fig:dark}. The  distribution provides a direct measurement of dark noise (DCR in Fig.~\ref{fig:dark}), direct cross talk (CT) and the occurrence of after-pulses (AP).    
The main group of events is due to primary, Poisson-distributed, dark counts that make up the DCR. The amplitude is centered at 18~mV for the CAB-based test setup. The value of the first photoelectron voltage peak, and the distribution in time is exponential with a decay time corresponding to the inverse of the DCR. The CT events occur at very short time after the preceding pulse, which is the time needed for the cross-talk photon to reach a neighboring cell and trigger an independent avalanche. The time scale of this effect is at ps level and therefore the time distribution overlaps with DCR pulses, but the pulses have larger amplitudes corresponding to two or more photoelectrons. Afterpulses occurs when, during an avalanche, an electron is trapped by an impurity in the lattice and is then released after a characteristic time, generating a second avalanche. Since the afterpulse event and its primary avalanche occur in the same cell, the time distribution is determined by both the traps time constants and the recharge time constant of the cell. When the time distance is lower than the full cell recharge, the resulting pulse has a reduced amplitude.
Afterpulses thus produce events separated in time
whose amplitude is 1PE for large delays (full recharge) or lower if the pulse occurs before the cell is fully recharged.

The overall size of correlated noise and the percentage of AP and CT, compared with the bulk of DCR, changes as a function of the overvoltage. CT and AP expressed as fraction of the DCR are tabulated in Tab.~\ref{tab:CTAP} for +3 and +4 OV. These measurement were recorded using the CAB-based setup in dark-room at 77~K.

\begin{table}
    \centering
    \begin{tabular}{|c|c|c|}
    \hline
    OV [V] & Crosstalk [\%] & Afterpulses [\%]  \\
    \hline\hline
    3 & 16.6 & 2.7\\
    \hline
    4 & 25.4 & 4.7\\
    \hline
    5 & 30.1 & 7.3\\
    \hline
    \end{tabular}
    \caption{Evolution of the fraction of crosstalk and afterpulse at T~=~77~K, in percentage of the DCR, as function of the overvoltage.}
    \label{tab:CTAP}
\end{table}

\subsection{Response of the tiles}
\label{sec:ganging}
The tile assembly is well suited to test ganging configurations of interest for implementations in the X-Arapuca. In particular, up to four NUV-HD-SF tiles (each with 6 SiPMs in parallel) were connected to amplifier input. Unlike the implementation of ProtoDUNE-S,P where the SiPMs are connected in parallel to a warm amplifier located in the front-end electronics~\cite{Abi:2017aow}, both in our setup and in DUNE the amplifier will be located in the proximity of the X-Arapuca. The response of the SiPMs connected in parallel and summed in one tile, together with the response of multiple tiles were investigated using the CAB of Fig.~\ref{fig:Board}. 

\begin{figure}[htbp]
    \centering
    \includegraphics[keepaspectratio=true,scale=0.45]{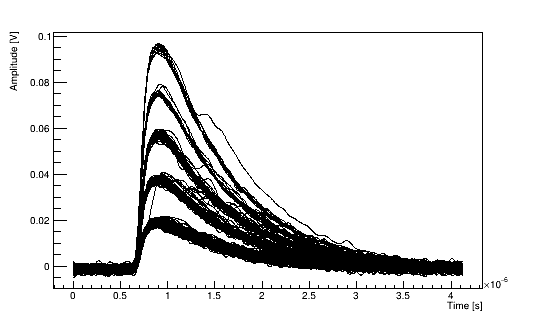}
    \caption{Waveforms acquired with SiPM bias, at 4~V of OV and T~=~77~K, triggered by the LED pulser.}
    \label{fig:waveform}
\end{figure}

For such analysis the SiPM were illuminated by the LED driver to an average photon intensity corresponding to $\sim$~2.5~p.e. (Fig.~\ref{fig:waveform}) per trigger. In each run, we recorded 10$^4$ events with the external trigger given by the LED pulser.
The SiPMs were studied as standalone devices or connected in parallel in groups of two and four SiPMs. Finally, we recorded runs connecting up to four tiles to the input of the cold amplifier. Since there are no series connections among SiPMs, the voltage at the power supply is the same as the bias voltage of the SiPMs. In these runs, the devices were all biased at the same voltage, +3~V and +4~V of overvoltage, depending on the run.

The SiPM network topology is expected to increase the series noise of the system due to the increase of the equivalent capacitance of the SiPMs. In these runs we studied the evolution of the signal-to-noise (S/N) ratio as a function of the number of SiPMs that were connected to the amplifier.  

\begin{figure}
    \centering
    \includegraphics[keepaspectratio=true,scale=0.4]{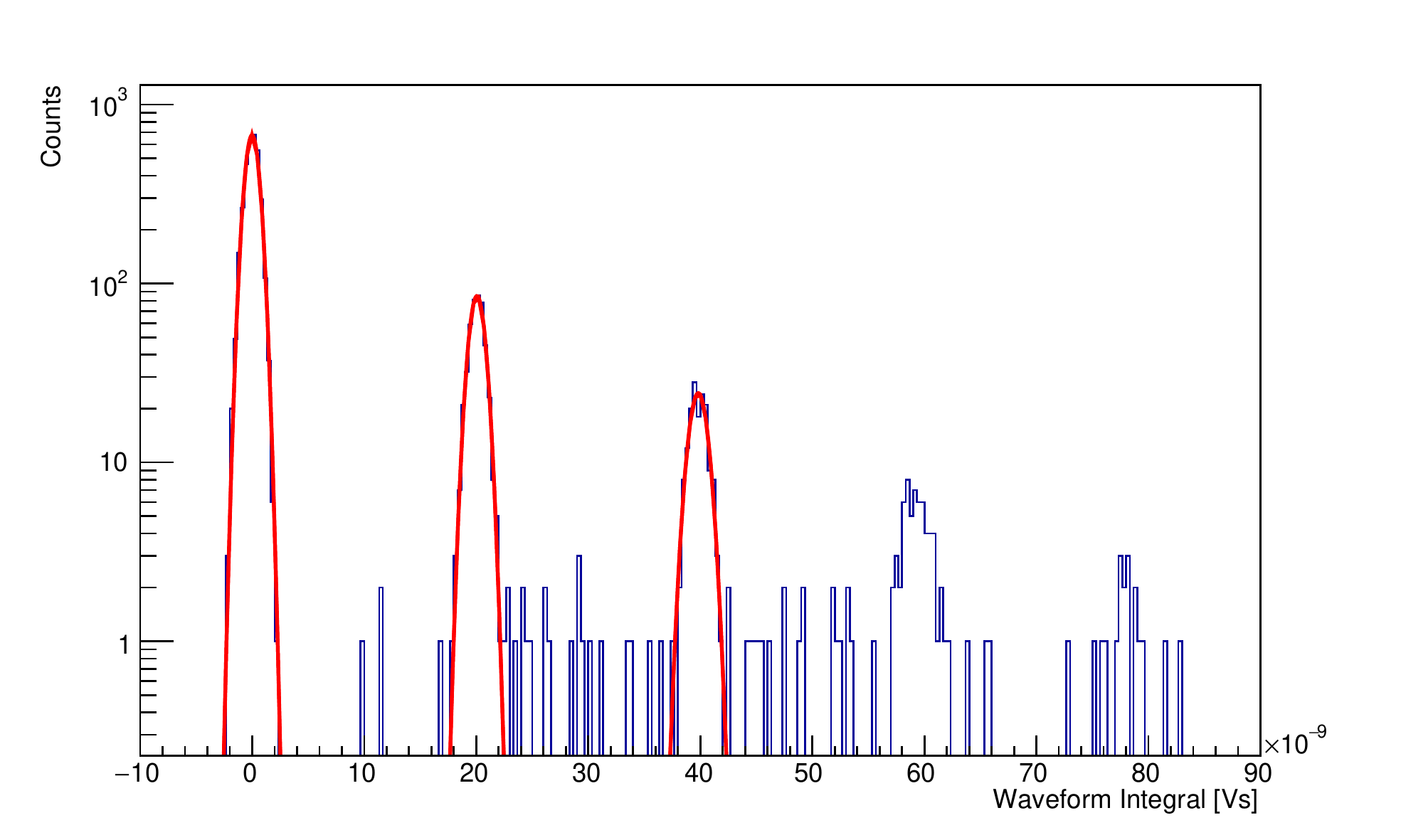}
    \includegraphics[keepaspectratio=true,scale=0.4]{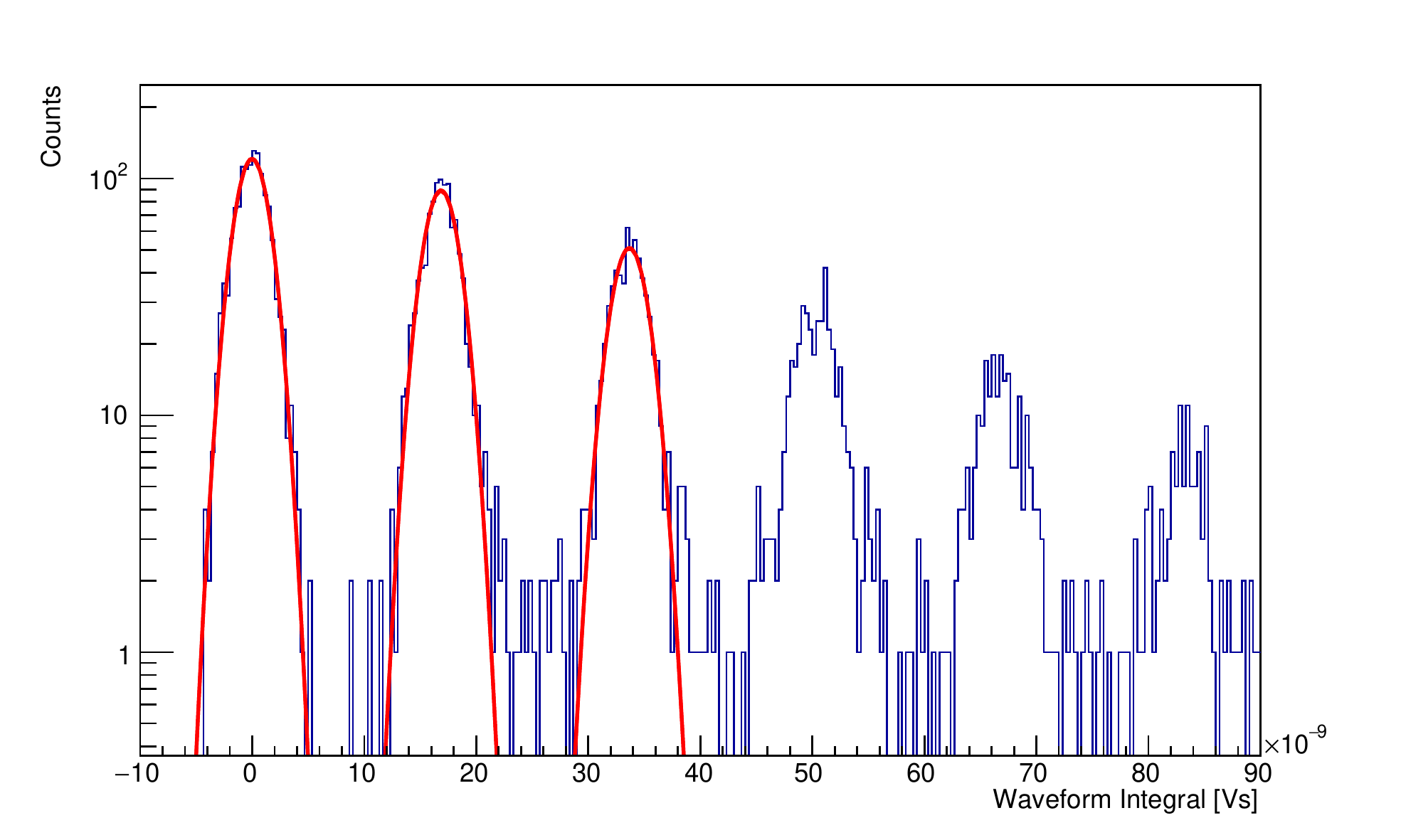}
    \caption{Integral of the SiPM waveforms, with one SiPM (top) ant four tile (bottom) in ganging, with bias at 4~OV and T~=~77~K.}
    \label{fig:integral}
\end{figure}

The S/N was computed from the distribution of the integral from 0 to 4~$\mu$ s of the signal waveforms (Fig.~\ref{fig:integral}). It is defined as the ratio between the distance of the first p.e. peak and the noise peak in the distribution of the integrals, divided by the $\sigma$ of the first p.e..
The means and the $\sigma$s are computed fitting with a Gaussian functions the distributions of the waveform integrals corresponding to noise and different p.e..
The gaussian fits are shown in Fig.~\ref{fig:integral} (red lines)
for the two most extreme cases: a single SiPM connected to the amplifier (top plot) and 4 tiles (24 SiPMs) connected in parallel (bottom plot).
Table \ref{tab:noise} summarizes the S/N as a function of the number of SiPMs. 
%Note in particular that the S/N increases with overvoltage up to +4 OV and decreases above this value due to the increase of the afterpulses. 

%From the distance between the peaks of the distribution the gain is identified, while the ratio between the mean and the width of the first photoelectron peak give the S/N.

%The gain results 0.8~x~10$^6$ per 1~V of overvoltage, both at T~=~77~K and at room temperature.

%The S/N is affected by the number of SIPM connected in parallel, since the noise sums up. With the active ganging the S/N is anyway preserved and remains >10 also with 4 tiles connected together (Tab.~\ref{tab:noise}).

\begin{table}
    \begin{tabular}{|c||c|c|c|c|c|c|c|c|}
    \hline
    \textbf{SiPM} & 1 & 2 & 4 & 6$\times$1 & 6$\times$2 & 6$\times$3 & 6$\times$4 \\ 
    \hline
    \textbf{mm$^2$} & 16 & 32 & 64 & 96 & 192 & 288 & 384 \\  
    \hline\hline
    3 OV & 26.6 & 21.3 & 21.3 & 21.4 & 15.1 & 13.1 & 11.9 \\
    \hline
    4 OV & 28.1 & 27.9 & 26.7 & 24.3 & 17.9 & 13.3 & 12.0 \\
    \hline
 %   5 OV & 25.8 & 23.7 & 21.5 & 20.4 & 12.3 & 9.7 &  \\
 %   \hline
    \end{tabular}
    \caption{S/N as a function of the OV and of the number of SiPMs connected, i.e. of the total active area, at T~=~77~K. 6~$\times$~n corresponds to n tiles.}
    \label{tab:noise}
\end{table}

Unlike the S/N, the rise time of the signal is mostly dominated by the bandwidth of the second-stage amplifier and does not depend on the number of SiPMs connected. The falling edge (90\%~$\rightarrow$~10\% of the peak amplitude) instead increases from 1.8~$\mu$s to 2.1~$\mu$s connecting up to 24 SiPMs (see Tab.~\ref{tab:rising-fall}).

\begin{table}
    \begin{tabular}{|c|c|c|}
    \hline
    Number of SiPMs & Rising time [ns] & Falling time [$\mu$s]\\
    \hline\hline
    1 & 120 & 1.85 \\ 
    \hline
    2 & 130 & 1.80  \\
    \hline
    4 & 126 & 1.87 \\
    \hline\hline
    Number of tiles & Rising time [ns] & Falling time [$\mu$s]\\
    \hline\hline
    1 & 129 & 1.91 \\ 
    \hline
    2 & 141 & 2.05  \\
    \hline
    3 & 134 & 2.09  \\
    \hline
    4 & 134 & 2.15 \\
    \hline
    \end{tabular}
    \caption{Rising and falling time (10\%-90\% of the peak) for different SiPM configurations biased at +4~OV, at T~=~77~K.}
    \label{tab:rising-fall}
\end{table}

\section{Conclusion}
Liquid argon experiments at multi-kton scale and, in particular, the DUNE 10~kton modules pose unprecedented challenges to the field of cryogenic solid state photodetectors. Even if in the last decade progress in this field was spectacular, cryo-reliability, reversibility of thermal and electrical properties, packaging and detector performance remain very active fields of research. In this paper, we addressed a technology that is particularly well suited for cryogenic applications. NUV-HD-SF sensors fulfill cryo-reliability requirements at the level of the silicon substrate, bonding and packaging. They achieve 200~mHz/mm$^2$ DCR at 77~K for a nominal overvoltage of +4~V with cross-talk and afterpulse at the level of 25\% and 4.7\%, respectively. Light collection efficiency in liquid argon detectors can be enhanced by optical traps as the DUNE X-ARAPUCA, which requires ganging of a large number of sensors in a single electronic channel. The 6 SiPM tile of 4~$\times$~4~mm$^2$ devices is a convenient form factor for these implementations. In particular, we demonstrated that cold amplification by a single TIA with a BJT at front-end provide a signal to noise of 28 for a single SiPM.  S/N is equal to 12 after ganging four tiles (24 sensors, 3.8~cm$^2$ of active surface). Further developments include the implementation of the low-noise, high dynamic range cold amplifier recently developed in~\cite{ampl}, an additional reduction of the DCR using the new NUV-HD-Cryo technology optimized for the X-ARAPUCA and ganging of sensors at the level of $\mathcal{O}(10)$~cm$^2$.     

\section*{Acknowledgement}
The authors gratefully acknowledge the electronics and mechanical division of INFN Bologna, Milano-Bicocca and Milano for support in the construction of the test facilities, and M.~Peracchi and F.~Romeo for help during the experimental campaign. This work was supported by the Dep. of Physics "G. Occhialini", Univ. of Milano-Bicocca (project 2018-CONT-0128).

\bibliographystyle{elsarticle-harv}

  \end{document}